\documentclass[hyper]{JHEP3}
\pdfoutput=1
\usepackage{epsfig}
\usepackage{latexsym}
\usepackage{amsfonts}
\usepackage{amsmath}
\usepackage{amsthm}
\usepackage{amssymb}
\usepackage{amsbsy}
\usepackage{multirow}
\usepackage{extarrows}

\newcommand{\CC}{\mathbb{C}} 
\newcommand{\RR}{\mathbb{R}} 
\newcommand{\ZZ}{\mathbb{Z}} 
\newcommand{\NN}{\mathbb{N}} 

\def\tr {{\rm tr}}

\def\calc         {{\cal C}}
\def\cald         {{\cal D}}

\def\calf         {{\cal F}}

\def\calm         {{\cal M}}

\def\calo         {{\cal O}}
\def\calp         {{\cal P}}

\def\calr         {{\cal R}}
\def\cals         {{\cal S}}
\def\calt         {{\cal T}}

\newsavebox{\uuunit}
\sbox{\uuunit}
    {\setlength{\unitlength}{0.825em}
     \begin{picture}(0.6,0.7)
        \thinlines
        \put(0,0){\line(1,0){0.5}}
        \put(0.15,0){\line(0,1){0.7}}
        \put(0.35,0){\line(0,1){0.8}}
       \multiput(0.3,0.8)(-0.04,-0.02){12}{\rule{0.5pt}{0.5pt}}
     \end {picture}}

\def\be{\begin{equation}}
\def\ee{\end{equation}}
\def\bea{\begin{eqnarray}}
\def\eea{\end{eqnarray}}

\def\a{\alpha}
\def\b{\beta}
\def\h{\eta}

\def\d{\delta}
\def\e{\epsilon}

\def\l{\lambda}
\def\L{\Lambda}

\def\f{\phi}

\def\m{\mu}
\def\n{\nu}
\def\o{\omega}
\def\O{\Omega}
\def\p{\pi}
\def\r{\rho}

\def\sF{{{ F}\!\!\!\!\hskip.8pt\hbox{\raise1pt\hbox{/}}\,}}
\def\som{{{ \omega}\!\!\!\!\hskip.8pt\hbox{\raise1pt\hbox{/}}\,}}
\def\sJ{{{\rm J}\!\!\!\!\hskip.8pt\hbox{\raise1pt\hbox{/}}\,}}

\def\pa{\partial}

\def\to{\rightarrow}
\def\nonu{\nonumber \\{}}
\def\half{{1 \over 2}}

\title{Quantization of conical spaces in 3D gravity}
\author{Joris Raeymaekers\\
Institute of Physics of the ASCR, \\ Na Slovance 2, 182 21 Prague 8, Czech Republic.\\
email: {\rm \texttt{joris@fzu.cz}}}

\abstract{We discuss the quantization and holographic aspects of a class of conical spaces in 2+1 dimensional pure AdS gravity.
 These appear as topological solitons   in the Chern-Simons formulation of the theory  and are closely related to
  the recently studied conical solutions in higher spin gravity. We discuss the classical fluctuations around these solutions, which form exceptional coadjoint orbits of the asymptotic Virasoro group.
 We argue that the quantization of these solutions leads to nonunitary representations of the Virasoro algebra, on  account of their having boundary graviton fluctuations which lower the energy.
 We propose a framework to quantize them in a semiclassical expansion in the inverse central charge, which we use to compute their one-loop corrected energies.  Interestingly, the  resulting Virasoro representations contain a null vector, thus providing an appearance of Kac's degenerate representations, which are nonunitary at large central charge,
 in the context of gravity. We  match the computed quantum corrections in the bulk with the properties of a class of primaries  in Kac's  classification.}

\begin{document}

\section{Introduction}
Pure gravity in 2+1 dimensions with negative cosmological constant   has  proven to be an interesting laboratory to test ideas in quantum gravity and holography. Despite having no local degrees of freedom, it possesses an asymptotic Virasoro symmetry \cite{Brown:1986nw} which can be viewed as the symmetry of a dual 1+1 dimensional CFT \cite{Maldacena:1997re}.  Furthermore the gravity theory contains black hole solutions \cite{Banados:1992wn} whose entropy can be understood from modular invariance of the dual CFT \cite{Strominger:1997eq}. Despite these qualitative successes, it is   still unclear whether pure  gravity at weak coupling (large central charge  $c$)  exists as a quantum theory   and, if so, what are the properties of the dual CFT \cite{Witten:2007kt},\cite{Gaberdiel:2007ve},\cite{Maloney:2007ud}.

An important step in identifying   a  candidate dual CFT is to understand  the spectrum  of smooth, asymptotically AdS, classical
solutions, which are believed to represent semiclassical  states in the spectrum of the dual CFT. In this note, we will illustrate a subtlety
in the concept of smoothness related to the  formulation  of the theory in terms of Chern-Simons gauge fields \cite{Achucarro:1987vz},\cite{Witten:1988hc}.  In this formulation the metric is a derived quantity and in particular there is no natural way to impose its invertibility. As a consequence, many observables that exist in the metric  formulation and which involve the inverse metric,
are not natural observables in the Chern-Simons formulation. Instead, the natural observables in the Chern-Simons formulation are based on  holonomies of the gauge field, and
we will see that these provide a somewhat cruder measure of smoothness.

The solutions we will study are labelled by an integer $s$  greater than one  and correspond to metrics with a conical singularity of excess  angle $2 \p (s-1)$.
Conical defects in 2+1 dimensional
 gravity have been
extensively studied following the pioneering work \cite{Deser:1983tn},\cite{Deser:1983nh}. The conical excess solutions were, as far as we know,  first explored in \cite{Izquierdo:1994jz} as BPS solutions of 3-dimensional supergravity, and were subsequently studied in  \cite{Mansson:2000sj},\cite{Balasubramanian:2000rt}. Despite being metrically singular, we will argue
that the conical excess solutions appear smooth to the natural observables in the Chern-Simons formulation.
 One new insight coming from our analysis is that requiring the solutions to appear smooth to the point particle probes considered in \cite{Witten:1989sx},\cite{Ammon:2013hba}  picks out the conical spaces without angular momentum.

Aside from  the singularity in the metric, another reason that conical excesses are usually discarded is that they have energies below that of the global AdS solution, implying that they cannot be part of the spectrum of
a unitary dual CFT. Nevertheless, many consistent nonunitary 2-dimensional CFTs are known to exist, some of which  have gravity duals\footnote{Known  nonunitary examples of holographic duality include the gravity/logarithmic CFT duality (see \cite{Grumiller:2013at} for a review and further references),  and the large $N$ gauge theories based on a supergroup which were recently discussed in \cite{Vafa:2014iua}.}, and one might expect the conical spaces to play a role in this context.   Indeed, the conical excesses are the pure gravity avatars  of similar solutions in  higher spin gravity theories\footnote{At first sight, the pure gravity  conical spaces may seem
 metrically more singular than their cousins in higher spin theories, since for the latter one can always apply a higher spin gauge gauge transformation such that the metric has no curvature singularity \cite{Castro:2011iw}. However we should keep in mind that metric curvature invariants are not good observables in higher spin theories; for example one can  construct
 higher spin solutions where the metric appears in some gauges to be regular yet the Chern-Simons gauge field describing them is singular.}, which were argued to have a consistent dual interpretation in a nonunitary semiclassical limit of the dual CFT\cite{Castro:2011iw},\cite{Perlmutter:2012ds},\cite{Hijano:2013fja},\cite{Campoleoni:2013iha}. The arguments leading to this identification were so far made in the classical approximation in the bulk, and the
 main motivation of this work was to develop a framework to include  bulk quantum corrections  in the simplified setting of pure  gravity.

We will see that the conical solutions in pure gravity are topological solitons characterized by a   winding number.
To discuss some of their quantum aspects we will follow the  standard approach for quantizing solitons, namely to quantize the fluctuations around the classical solution (see \cite{Coleman} for a review and further references). In the case of pure three dimensional gravity,  the fluctuations around a given solution come from acting on it with  asymptotic Virasoro generators, or in other words  from dressing the solution with boundary graviton excitations. 
 Exponentiating these infinitesimal variations to yield an action of the Virasoro group on the  solution, one  obtains what is called a coadjoint orbit of the Virasoro group. Each orbit carries a natural
 Poisson bracket,  which upon quantization should lead to a Virasoro  representation. The Virasoro coadjoint orbits were classified in \cite{Lazutkin},\cite{Segal:1981ap} and further analyzed by Witten \cite{Witten:1987ty}. Subsequent work includes \cite{Alekseev:1988ce},\cite{Bershadsky:1989mf},\cite{Balog:1997zz}, see  also \cite{Garbarz:2014kaa},\cite{Barnich:2014zoa} for  recent discussions of coadjoint orbits in the context of three dimensional gravity.

 It turns out that the conical spaces give rise to rather special orbits: they are the `exceptional' orbits which possess an $SO(2,1)^{(s)} \times SO(2,1)^{(s)}$
symmetry. Here, the superscript $s$ indicates  that the  group is an $s$-fold cover of $SO(2,1)$. The generators of this  symmetry  are embedded in the Virasoro algebra in a different manner for each conical space and, in particular, are different  from the symmetry generators of  the global AdS solution.
 By analyzing the  exceptional coadjoint orbits  one finds that the conical spaces possess boundary graviton fluctuations which lower the Virasoro energy, which is unbounded below \cite{Witten:1987ty}.
For this reason  they cannot be quantized in a way that leads to a unitary highest weight Virasoro representation, and it has so far proved impossible
to quantize the exceptional orbits by standard methods.

In this work we will propose a framework  to quantize the exceptional orbits which leads to a nonunitary highest  weight
  representation of the Virasoro algebra,  constructed as a  semiclassical perturbation expansion in the inverse central charge,
    and proceed to  compute the 1-loop correction to the energy of the conical spaces.
  Furthermore, as anticipated in \cite{Witten:1987ty},  the quantized exceptional orbits contain a null vector at level $s$. Hence the conical spaces constitute an appearance of the most interesting representations of the Virasoro algebra, the degenerate ones, in a gravity context. Our results for the energy correction and the null vector lead us to identify the
 quantized conical spaces with the degenerate representations of the type $(1,s)$ in Kac's classification \cite{Kac:1978ge}, which are well-known  to be nonunitary at large values of the central charge $c$. It is fascinating that these representations become unitary, and belong of the spectrum of the Virasoro minimal models, at small
 values of the central charge $0<c<1$, which represents the strong coupling regime of the gravity theory.

\section{Chern-Simons formulation of 3D AdS gravity}\label{secCS}
Let us  briefly review the Chern-Simons formulation of gravity and state our conventions.
 Pure gravity in 2+1 dimensions with negative cosmological constant can be reformulated as a Chern-Simons theory with gauge group $SO(1,2)\times {SO(1,2)}$ with opposite levels for the two factors \cite{Achucarro:1987vz},\cite{Witten:1988hc}:
\bea
S &=& S_{CS}[A] - S_{CS}[\tilde A]\\
S_{CS}[A] &=& {k \over 4 \p} \int_\calm \tr_3 \left( A\wedge d A + { 2\over 3} A\wedge A \wedge A \right).\label{SCS}
\eea
The gauge potentials $A, \tilde A$ take values in the $so(1,2)$ Lie algebra whose generators $J_a, a = 0,1,2$ satisfy the commutation relations\footnote{Indices are raised with the metric $\h_{ab} = diag (-1,1,1)$ and $\e_{012} \equiv 1$.}
\be
[ J_a, J_b] = \e_{ab}^{\ \ c} J_c\label{Js}
\ee
 Viewing $SO(1,2)$ as the 2+1-dimensional Lorentz group, the generator $J_0$ is compact and corresponds to spatial rotations, while $J_{1,2}$ generate boosts.
In our conventions the trace in (\ref{SCS}) is taken in the defining three dimensional representation\footnote{Some useful properties are
$\tr_3 (J_a J_b) = 2 \h_{ab},
\tr_3 (J_a J_b J_c) = \e_{abc}$.}.
The vielbein and spin connection are obtained from
\bea
A &=& \left( \o^a + {e^a \over l}\right) J_a\nonu
\tilde A &=&  \left( \o^a - {e^a \over l}\right)  J_a
\eea
where $\o^a=\half \e^a_{\ bc} \o^{bc}$ and $l$ is the AdS radius.

The Chern-Simons field strengths are related to the torsion $\calt^a =  d e^a +\e^a_{\ bc} \o^b \wedge e^c $
and curvature $\calr^a= d \o^a + \half \e^a_{\ bc} \o^b \wedge \o^c$ two-forms    as follows:
\bea
{l\over 2} (F- \tilde F ) &=& \calt\\
\half( F + \tilde F)  &=&\calr + {1\over  l^2} \e \wedge e.
\eea
Hence the equations of motion, which impose the flatness  of $A$ and $\tilde A$, imply that the connection $\o$  is torsionless and that Einstein's equations hold with negative cosmological constant $\L = -{2\over l^2}$.

Writing the action in terms of $e, \o$ gives
\bea
S &=& {k \over  \p l}\int_\calm d^3 x \sqrt{-g} \left( R + { 2\over l^2} \right) +{k \over  \p l}\int_{\pa \calm} e^a \wedge \o_a .
\eea
This allows us to  obtain Newton's constant from
\be
k = {l \over 16 G}.
\ee
The Brown-Henneaux central charge \cite{Brown:1986nw} is the combination
\be
c = {3 l \over 2 G}= 24 k.
\ee
The large  $c$ (weak coupling) limit is the semiclassical regime, where the path integral is dominated by  classical gravity solutions,
while for small $c$ (strong coupling) quantum corrections are significant.

Gauge transformations act as
\bea
A \to \L^{-1} A \L + \L^{-1} d\L\nonu
\tilde A \to \tilde \L^{-1} \tilde A \tilde \L + \tilde \L^{-1} d\tilde \L\label{gaugetransf}
\eea
and in order for  $e^{i S}$ to be gauge-invariant, $k$ must be quantized in integer units \cite{Witten:2007kt}, which implies that $c$   is a multiple of 24\footnote{ As explained in \cite{Witten:2007kt}, if we replace the gauge group by an $n$-fold cover of $SO(1,2)$,
$k$ is quantized in units of $n^{-2}$ and $c$ is a multiple of $24/n^2$. For example taking the gauge group to be $SL(2, \RR)$, $c$ should be a multiple of $6$.}. By contrast, if we  consider   Euclidean gravity, the  relevant  Chern-Simons gauge group is replaced by $SL(2,\CC )$ and there is no quantization condition  on $k$ and $c$ \cite{Witten:1989ip} (this is one of the reasons why the relation between Lorentzian and Euclidean Chern-Simons gravity is poorly understood).     For most of this paper we will consider the Lorentzian theory but will occasionally comment on the Euclidean case.

\section{Asymptotic symmetries and coadjoint orbits}\label{secas}
In this section we will review the analysis of the asymptotic symmetries in the Chern-Simons formulation \cite{Banados:1998gg} and  rephrase these standard results in the language of coadjoint orbits of the Virasoro group, which we will use in section \ref{secquant}.
We start by using the isomorphism $so(1,2) \sim sl(2,\RR )$ to introduce a new Lie algebra basis $\{ V_0 , V_{\pm 1} \}$
\bea
V_0 &=& - J_2\nonu
V_{\pm 1} &=& J_0 \pm J_1.
\eea
which obey $sl(2,\RR )$ commutation relations
\be
[V_m, V_n ] = (m-n) V_{m+n}.
\ee
Note that in the new basis, the compact generator is $\half ( V_1 + V_{-1} )$.
We will take $\calm$ to have the same topology as the global AdS$_3$ manifold, namely that of the solid cylinder $\RR \times D$. The time  coordinate $T$ runs along the length of the cylinder while on the disk $D$ we choose polar coordinates $\r, \f$, where $\f$ has period $2\p$.

Gauge connections $A, \tilde A$  satisfying asymptotically  AdS boundary conditions can be  gauge-fixed to the form \cite{Banados:1998gg}
\bea
A &=& g^{-1} a(x^+) g dx^+ + g^{-1} d g, \qquad g = e^{\r V_0}\nonu
\tilde A &=&  g \tilde a(x^-)  g^{-1} dx^- +  g d  g^{-1}\label{partialgf}
\eea
where $x^\pm = \f \pm T$ and $a(x^+), \tilde a(x^-)$ are Lie algebra valued functions which are assumed to be in the so-called highest weight gauge
\bea
a &=& V_1 -  {12\p \over c} t(x^+) V_{-1}\nonu
\tilde a &=& V_{-1} - {12\p \over c} \tilde t(x^-)  V_{1}.\label{hwgauge}
\eea
The functions $t (x^+), \tilde t (x^- )$ parameterize the phase space of flat connections obeying asymptotically AdS boundary conditions. As we will see below, they are to be identified with
the left- and right-moving boundary stress tensors.
The global AdS$_3$ solution corresponds to taking
\be t_{AdS}= \tilde t_{AdS}= - {c \over 48\p}. \label{adscov}\ee

At fixed time, say  $T=0$, the form  (\ref{partialgf}) is preserved by gauge parameters of the form
\be
\l = g^{-1} \e (\f) g, \qquad \tilde \l= g \tilde \e(\f) g^{-1}.
\ee
under which $a, \tilde a$ transform as
\be
\d a = \e' + [a, \e], \qquad  \d \tilde a = \tilde \e' + [\tilde a, \tilde \e].
\ee
 Decomposing $\e, \tilde \e$ as
\bea
\e (\f) &=& j (\f)V_1 + \e_0 (\f) V_0+\e_{-1} (\f) V_{-1}\nonu
\tilde \e (\f) &=& \tilde j (\f) V_{-1} + \tilde \e_0 (\f ) V_0+\tilde \e_{1} (\f )  V_{1}\label{asgauge}
\eea
the requirement that the gauge transformations  preserve the highest weight gauge (\ref{hwgauge}) fixes $\e_0, \e_{-1}$ and $\tilde \e_0, \tilde \e_{1}$  in terms of $j$ and $\tilde j$ respectively. Such transformations should be viewed as infinitesimal asymptotic symmetries, which are therefore parameterized by  two functions on the circle $j(\f )$ and $\tilde j (\f )$. Under the  action of the infinitesimal asymptotic symmetries  $t(\f)$ and   $\tilde t(\f)$ transform as
\bea
\d_j t &=& 2 t  j ' + j t' - {c \over 24 \p } j'''\nonu
\d_{\tilde j} \tilde t &=& 2\tilde t \tilde j ' + \tilde j \tilde t' - {c \over 24 \p} \tilde j''' \label{transfo}
\eea
The conserved charge $q_j$ corresponding to the asymptotic symmetry $j(\f)$ is
\bea
q_j(t) &=& \int_0^{2\p} d \f j (\f)  t (\f)\label{Noethercharges}
\eea
and similarly for the right-moving charges $\tilde q_{\tilde j}$ (for the rest of this section, we will display only the formulas in the left-moving
sector, the right-moving side proceeding analogously). The  conserved charges are linear functionals on the phase space and we would like to compute their Poisson
brackets. These can be deduced from the fact for every function $\calo$ on the phase space, its variation under $j$ arises from the Poisson bracket with $q_j$: $ \{q_j, \calo \}_{PB}= \d_j \calo$. Taking $\calo$ to be the conserved charge  $\calo = q_k$ we find the Poisson bracket
\bea
\{ q_j, q_k \}_{PB} (t) &=& \int_0^{2 \p} d\f k \d_j t\\
 &=& q_{(j'k - j k')}(t) - {c \over 48 \p} \int_0^{2 \p} d\f (j''' k - j k''')\label{PBVir}
 \eea

 Let us review the group theoretic meaning of this expression.
 If $c$ were zero, this Poisson bracket would realize the Lie algebra $diff (S^1 )$ of reparametrizations of the circle in the sense
 that
 \be
  \{ q_j, q_k \}_{PB} = q_{[k,j]}
  \ee
  where $[k,j] = j'k - jk'$ is the commutator in $diff (S^1 )$. Therefore $j(\f), k(\f) $ should be thought of as
  components of tangent vectors $j(\f) \pa_\f,\  k(\f) \pa_\f$. As for $t(\f)$, the expression for the charge (\ref{Noethercharges}) allows us to identify   $t(\f)$ as an element of the vector space dual  to $diff (S^1 )$.
 The transformation law (\ref{transfo}) at $c=0$ defines the coadjoint representation of $diff(S^1 )$ and shows that $t ( \f)$ should be seen as the component of a  quadratic differential $t(\f) d\f^2$.

  When $c$ is nonzero, the Poisson bracket (\ref{PBVir}) instead realizes a central extension of $diff (S^1 )$, the Virasoro algebra,  as we shall presently review.
  Extending $diff (S^1 )$ with a central generator  $\hat c$, elements  of the   Virasoro algebra can be represented as pairs
  \be (j(\f), n) \longleftrightarrow j (\f) \pa_\f + n \hat c \ee
where  $n$ is a real number. The commutation relations are
\be
[(j , n ), (k , m )  ]= \left( (j k' - j'  k ), {1 \over 48 \p} \int_0^{2\p} d\f (j'''  k - j   k''' )\right).
\ee
Similarly, we extend the dual vector space by including the constant parameter $c$  as an extra  coordinate. The dual vector space to the Virasoro  algebra consists of pairs  of the form $(t(\f ),c)$, and the pairing between adjoint and coadjoint vectors is given by the generalization of the expression (\ref{Noethercharges}) for the charge:
\be
q_{(j,n)} (t,c)  =  \int_0^{2\p} d \f j   t  + c n \equiv \langle (t,c), (j,n)\rangle \label{pairing}
\ee
The  transformation law (\ref{transfo}) can be extended in such a way that the pairing is Virasoro-invariant, meaning
\be\langle \d_{(k,m)} (t,c), (j,n)\rangle + \langle  (t,c), [(k,m), (j,n)]\rangle=0 .\ee
This leads to
\be
\d_{(j,n)}( t, c) = ( 2 t  j ' + j t' - {c \over 24 \p } j''',0).\label{coadjinf}
\ee
This transformation law defines the coadjoint representation of the Virasoro algebra.  Using these definitions one finds that the Poisson bracket (\ref{PBVir}) can be written as
\be
 \{ q_{(j,n)}, q_{(k,m)} \}_{PB} (t,c) = q_{[(k,m),(j,n)]} (t,c)\label{KK}
 \ee
from which we see that the Poisson bracket (\ref{PBVir}) indeed realizes the Virasoro algebra.

The Poisson bracket  can be written in a more standard form by choosing the following basis  for the charges  generating $diff(S^1 )$:
\be
l_m = q_{(e^{i m \f},0)}.
\ee
In this basis the Poisson brackets take the form
\be
-i \{ l_m, l_n \}_{PB} =(m-n)l_{m+n} +{\hat c m^3 \over 12} \d_{m,-n}.\label{VirPB}
\ee
 Note that our Virasoro energies $l_0, \tilde l_0$ are naturally defined on the boundary cylinder, and on the global  AdS solution (which will turn out to correspond  to the $SL(2,\RR )$ invariant  vacuum in the dual CFT) they take the values
\be
l_0 ((t_{AdS},c)) = \tilde l_0 ((\tilde t_{AdS},c)) = -{c \over 24}.
\ee
This corresponds  to the Casimir energy of a cylinder of circumference  $2 \p$. To make contact with the more standard conventions where $l_0$ and $\tilde l_0$ act on the plane one should shift
them by ${c \over 24}$.

 Extending this analysis to include the right-moving sector, we conclude that the asymptotic charges generate two copies of the Virasoro algebra through Poisson brackets. The
 combinations $(t(\f), c)$ and $(\tilde t(\f), c)$ transform in the coadjoint representation of the respective Virasoro algebras.
At fixed central charge $c$, asymptotically AdS solutions come in families obtained by acting on a given solution with the Virasoro symmetries, which are  are referred to as coadjoint orbits of the Virasoro group. Physically, moving around on a coadjoint orbit can be seen as dressing (or undressing) a given solution with boundary graviton excitations.
 A standard result, which we will use to quantize  coadjoint orbits in section \ref{secquant}, is that each  orbit possesses a natural symplectic form and hence a Poisson bracket (the Kirillov-Kostant bracket) such that the restrictions of the charges
to the orbit obey  (\ref{KK}).

Of course, not all coadjoint orbits are physically acceptable, as some of them may correspond to singular gravity solutions, and  the  issue of regularity will be the subject of the next section. The regular coadjoint orbits are expected to correspond to semiclassical states in the quantum theory, and upon quantizing them we expect
to obtain Virasoro representations belonging to  the spectrum of the dual CFT.

\section{Conical spaces as topological  Chern-Simons solitons}\label{seccon}
In this section we will revisit, in the pure gravity context, the conical solutions which were recently studied in higher spin gravity \cite{Castro:2011iw}. We will emphasize their similarity to solitons in the sense that they carry a topological winding number, and discuss their smoothness as seen by observables both in the metric and Chern-Simons formulations.
\subsection{Winding numbers}
Let us study the space of asymptotically AdS solutions on the solid cylinder in more detail.
At any fixed time $T$, the flat connections $a, \tilde a$ (see (\ref{hwgauge})) on the boundary cylinder can be expressed as
 \be a (\f) d \f = h^{-1} d h, \qquad  \tilde a (\f) d \f = \tilde h^{-1} d \tilde  h\label{puregauge}\ee
where  $h(\f), \tilde h (\f)$ are maps from the boundary circle into the gauge group $SO(1,2)$. These are
 classified by the first homotopy group  of  $SO(1,2)$, and since $SO(1,2)$ is contractible to its maximal compact subgroup $U(1)$, the first homotopy group is $\p_1 \left(SO(1,2)\right)= \ZZ$.
 Hence the  space of flat connections on the boundary  consists of topological sectors labelled by the two winding numbers $s, \tilde s$ of the maps $h (\f ), \tilde h (\f )$, which measure how many times the $U(1)$ directions in $SO(1,2) \times SO(1,2)$ are traversed  when we go around the boundary circle.
The following are representative maps with winding numbers $s,\tilde s$
\bea
m_s (\f) &=& e^{{s\f \over 2} (V_{1} + V_{-1})} \nonu
\tilde m_{\tilde s} (\f) &=& e^{{\tilde s\f \over 2} (\tilde V_{-1} +  \tilde V_{1})}.
\eea
These however don't give rise, upon substituting in (\ref{puregauge}), to  connections in the highest weight gauge  (\ref{hwgauge}).
This can be remedied by choosing different representatives  related by a constant gauge transformation
\bea
h_s (\f) &=& b\,  m_s (\f)\, b^{-1} = e^{ \f ( V_1 +{s^2 \over 4} V_{-1})}, \qquad b = e^{\ln {s \over 2 } V_0}\nonu
\tilde h_{\tilde s} (\f ) &=&  \tilde  b \,  m_{\tilde s} (\f)\, \tilde b^{-1} =  e^{ \f (  V_{-1} +{\tilde s^2 \over 4}  V_{1})},\qquad \tilde b = e^{- \ln {\tilde s \over 2 } V_0}.\label{hs}
\eea
Substituting in (\ref{puregauge}) gives  the following highest weight gauge connections\footnote{Here we encounter the subtlety that (\ref{hs}) is a good gauge transformation only when both $s$ and $\tilde s$ are positive. Since we can flip the sign of both $s$ and $\tilde s$ by sending $\f \to - \f$, we can only transform the maps with $s \tilde s >0$ to the highest weight gauge. We will assume from now on that  $s$ and $\tilde s$ are positive.}
\bea
a_s &=&  V_1 +{s^2 \over 4} V_{-1}\nonu
\tilde a_{\tilde s} &=&  V_{-1} +{\tilde s^2 \over 4} V_{1}.
\eea
We will label  these solutions  by their winding numbers $(s, \tilde s)$  in what follows.
The $(s, \tilde s)$ solution is characterized by  constant covectors $(t_s,c)$ and $(\tilde t_{\tilde s},c)$ with
\be t_s = - {c s^2 \over 48 \p }, \qquad  \tilde t_{\tilde s} = - {c \tilde s^2 \over 48 \p }.\ee
From (\ref{pairing}) we read off the Virasoro energies
\be
l_0 ((t_s,c)) = - {c s^2 \over 24}, \qquad
\tilde l_0 ((\tilde t_{\tilde s}),c) = - {c \tilde s^2 \over 24}.
\ee
The other Virasoro charges vanish on these solutions.
Comparing to (\ref{adscov}), we see that the case $s = \tilde s =1$ corresponds to global AdS: it is a somewhat peculiar feature of the Chern-Simons description
 that the  natural classical vacuum  of the theory appears in a winding sector.
 We note that the winding states with $s,\tilde s>1$  have energies below the AdS vacuum energy. Since in a unitary CFT all primaries have conformal weights above the $SL(2,\RR )$ invariant vacuum,  this is already an indication that these winding states can only play a role in nonunitary versions of holography,
as we shall see in more detail below.

\subsection{Metric-like observables}
Let us now discuss whether the $(s, \tilde s)$ winding solutions are smooth.
Since the $\f$-circles (i.e. the curves of constant $T, \r$) are by assumption contractible, our coordinate system is singular at some value of $\r$ (which we call the `origin'), where
the $\f$ coordinate is ill-defined. Our solutions satisfy the equations of motion everywhere except possibly in the origin.
To decide whether the solution is singular in the origin, we will look at suitable gauge-invariant observables which could measure such a singularity.
As we already mentioned in the Introduction, the analysis depends on whether we work in the Chern-Simons formulation, where the natural observables come from holonomies of the gauge field, or the metric formulation, where we have at our disposal the standard curvature invariants which are constructed using the inverse metric. Let's start by addressing smoothness in the  metric formulation.

 For simplicity  we restrict our attention to the solutions  where $\tilde s =s$ which have vanishing angular momentum\footnote{When $s \neq \tilde s$, the metric is that of a spinning conical defect:
$$
ds^2_{(s,\tilde s)} = l^2 \left[ d\r^2 + {s \tilde s\over 2} \cosh 2\r dx^+ dx^- - {s^2 \over 4 } (dx^+)^2 - {\tilde s^2 \over 4} (dx^-)^2\right]
$$
which apart from a curvature singularity also contains closed timelike curves.}.     The corresponding metrics are, after a shift
$ \r \to \r +  \ln { s\ \over 2}$,
\be
ds^2_{(s, s)} = l^2 \left[- s^2 \cosh^2  \r dT^2 +   d\r^2 + s^2 \sinh^2 \r d\f^2 \right]\label{staticdefs}
\ee
As anticipated in (\ref{adscov}), for $s =1$ we recover the standard global AdS metric. In the limit $s \to 0$ (where we cannot perform  the shift of the $\r$-variable) the metric  is that of the zero mass, zero angular momentum BTZ black hole
\be
ds^2_{(0,0)} = l^2 \left[ d\r^2 + e^{2 \r} \left(-dt^2 + d\f^2 \right) \right]
\ee
which has energy above that of global AdS.

For $s>1$ the metric (\ref{staticdefs}) can be seen to have a curvature singularity in $\r=0$, corresponding to a conical singularity with an excess angle of $2 \p (s-1)$ \cite{Deser:1983tn},\cite{Deser:1983nh}.
Let's rederive this fact in a way that emphasizes the difference between the metric and Chern-Simons formulations. In the metric formulation we postulate that the vielbein
is invertible.  If this is the case, we can construct many gauge-invariant observables besides the eigenvalues of the holonomies (\ref{hols}). For example, for every (non-null) two-surface we can define a surface observable
\be
\calo (\cals) = \int_\cals (F^a + \tilde F^a) e_a^\m n_\m\label{surfaceobs}
\ee
where $n_\m$ is the unit normal to the surface. 
Now let's evaluate $\calo (\cald)$ with $\cald$
a disc $0 \leq \r \leq \r_0$ at constant $T$, in the  background of the $(s,s)$ solution. For any solution with constant $t = \tilde t$ we can rewrite $\calo (\cald) $ as
\be
\calo (\cald) =   \int_\cald \left(\tilde R + {2\over l^2}\right) \sqrt{\tilde g} d\r d\f\label{integr}
\ee
with $\tilde g_{\m\n}$ the induced metric on $\cald$ and $\tilde R$ its scalar curvature.
Due to the equations of motion the  integrand vanishes everywhere, except possibly in the origin, and  we can replace (\ref{integr})  by
\be
\calo (\cald) =   \int_{\cald_\e}  \sqrt{\tilde g}  \tilde R d\r d\f
\ee
where ${\cald_\e} $ is a tiny disc around the origin.
The spatial metric near the origin is
\be
d \tilde s^2 \sim l^2 \left[ d\r^2 + s^2 \r^2 d\f^2 \right]
\ee
which can be written as the flat metric $l^2 du d\bar u$ in terms of the complex coordinate $u = \r e^{i s \f}$. However, the transformation from $(\r,\f)$ and $(u, \bar u)$ is not one-to-one, rather it is one-to-one for the coordinate $v = u^{1/s}$, in terms of which the metric is only conformally flat, with a conformal factor which is singular in the origin:
\be
d \tilde s^2 \sim l^2 |v|^{2(s-1)} dvd\bar v.
\ee
Hence a useful way to picture  the geometry near the origin   is as the Riemann surface of the function $u^{1/s}$, i.e. as an $s$-sheeted branched covering over the complex plane.
We then evaluate
\be
\sqrt{\tilde g} \tilde R =  (1- s)\pa_v\pa_{\bar v} \ln |v|^2 = 4\p (1-s) \d^{(2)}(v,\bar v)
\ee
from which we find
\be
\calo (\cald) = 4\p (1-s).
\ee
Hence from the metric point of view only the global AdS solution with $s=1$ is regular while the remaining winding states  are not.  The surface observable
 $\calo (\cald)$ which sees the singularity is not a natural observable in the Chern-Simons formulation of the theory as it explicitly  involves the inverse vielbein.
From the Chern-Simons point of view, the natural observables are rather based on the holonomies of the gauge field which we shall now discuss.

\subsection{Holonomies}

The natural observables in a three-dimensional Chern-Simons theory    are related to the holonomies of the gauge fields around a closed curve $\calc$:
\be
H ( \calc ) = \calp e^{\int_\calc  A}, \qquad
\tilde H ( \calc ) = \calp e^{\int_\calc \tilde A}\label{hols}
\ee
The eigenvalues of $H ( \calc ) , \tilde H ( \calc ) $ are gauge-invariant and, when the connections $A, \tilde A$ are flat,  depend only on the homotopy class of $\calc$.  However, they do depend on the representation
of the gauge group used to evaluate them.
 Since we are considering the pure gravity theory, which contains only the gauge fields in the adjoint representation, it's natural to evaluate the holonomies (\ref{hols}) in the adjoint representation, in our case the 3-dimensional representation of $SO(1,2)$.  For a smooth flat gauge connection, the  holonomy around a contractible loop should be unity. Here we shall take $\calc$ to be a  $\f$-circle, which is by assumption contractible,  in the    $(s, \tilde s)$ winding solution. We find that
\bea
H &\sim& e^{2 \p a} \sim e^{- 2 \p i s V_0}\nonu
\tilde H &\sim& e^{2 \p \tilde  a} \sim e^{- 2 \p i \tilde  s V_0}\label{holcon}
\eea
 where in the last step we have used
 \be
 \half ( V_1 + V_{-1} ) = M^{-1} ( - i V_0 ) M
 \ee
 with $M = e^{- i \p/4  ( V_1 - V_{-1} )}$. Since the eigenvalues of $V_0$ in the adjoint representation are $1,0$ and $-1$, we see that the holonomy is indeed trivial and  all the $(s, \tilde s)$ winding solutions appear to be  smooth \cite{Mansson:2000sj}. Note that, if we replace the adjoint representation with
 any other finite-dimensional representation of $SO(2,1)$, the holonomy remains trivial as the eigenvalues of $V_0$ are integer in these representations. Hence from the point of view of these observables, all the $(s, \tilde s)$ solutions appear to be regular.

More generally, we can ask if the $(s, \tilde s)$ winding states still appear smooth if we probe them with  external matter,
 for example a spinning point particle probe  of mass $M$ and spin $J$.
 It was argued in \cite{Witten:1989sx}  that such probes are related to  holonomies (\ref{hols}) evaluated in infinite-dimensional unitary representations of the gauge group. We shall follow here the recent discussion\footnote{See also \cite{deBoer:2013vca} for a closely related description of point particles in the Chern-Simons formulation, and \cite{Castro:2014mza} for the precise relation between the two descriptions.}  \cite{Ammon:2013hba} (see \cite{Castro:2014tta} for the extension to spinning particles) to which we refer for more details.
We will focus on the gauge-invariant Wilson loop operator
\be
W_R (\calc ) = {\rm tr}_R \left( \calp e^{\int_\calc ( A + \tilde A ) }\right).\label{Wline}
\ee
where $\calc$ is a closed loop and $R$ is a representation of the gauge group.
We  take   $R$ to be the infinite-dimensional highest weight representation of $SO(1,2)\times SO(1,2)$ built on a primary state $| h, \tilde h\rangle$:
\bea
L_1 | h, \tilde h\rangle &=& \tilde L_1 | h, \tilde h\rangle=0\nonu
L_0 | h, \tilde h\rangle &=& h | h, \tilde h\rangle, \qquad \tilde L_0 | h, \tilde h\rangle= \tilde h | h, \tilde h\rangle
\eea
The quantum numbers $h, \tilde h$ are related to the mass and spin of the probe as
\be
l M = h + \tilde h, \qquad J = h - \tilde h.
\ee
The physical meaning of this Wilson loop was found in \cite{Castro:2014mza} to be as follows. When $\calc$ is  noncontractible, the Wilson loop is related to the proper distance along  the curve (or a  generalization thereof for spinning particles). For example, for $J=0$ and $\calc$  the  $\f$-circle in the BTZ black hole background, $W_R (\calc )$ measures the Bekenstein-Hawking entropy. When $\calc$ is contractible, it measures rather the phase picked up by the wavefunction of the test particle\footnote{An argument for this goes as follows. Let's take $\calc$ to be a $\f$-circle and consider the
wavefunction of a position eigenstate, i.e. the
 propagator $\langle \r,\f, T|, \r',\f', T'\rangle$, which  can be represented as a sum over paths weighted by  (\ref{Wline}). When sending $\f \to \f + 2 \p$, each contributing  path $\calp$ gets deformed to a  new path  homotopic to $\calp + \calc$. Hence the phase picked up by the propagator is given by the Wilson loop around $\calc$.} when going around $\calc$, and when this phase is nontrivial it measures a singularity in the gauge field as seen by the probe. For example, for the $\f$-circle in the global AdS background one finds a trivial phase, while for conical defect solutions with $-c/48 \p< t, \tilde t <0$
 the phase is nontrivial, signaling a singularity.

 Let us now compute the Wilson loop in the background of our conical excess solutions labelled by $(s, \tilde s)$, where we take $\calc$ to be a $\f$-circle. The Wilson loop can be evaluated by representing the trace as a path integral over auxiliary variables, see  Appendix E in \cite{Castro:2014tta} for details in the case of spinning particles. For large $h, \tilde h$, where the point particle approximation is valid, the auxiliary path integral can be approximated by a saddle point contribution with the result
\be
W_R (\calc ) = e^{-\p  {\rm tr}_3\left(  (h \l_\f - \tilde h \tilde \l_\f ) V_0 \right) }\label{Wloopspin}
\ee
 where the trace is taken in the 3-dimensional representation of $SO(1,2)$.
The matrices $\l_\f, \tilde \l_\f$  are the eigenvalue matrices of $a_\f$ and $\tilde a_\f$ respectively, which we already determined in  (\ref{holcon}):
  \be
  \l_\f = -  i s V_0, \qquad  \tilde \l_\f = -  i \tilde s V_0.
  \ee
Substituting in (\ref{Wloopspin}) we obtain
\be
\boxed{W_R (\calc ) = e^{i \p \left( (s- \tilde s) M + (s + \tilde s) J \right)}.\label{phase}}
\ee
Let us discuss this result first for the case where the probe particles are bosons, so that $J$ is an integer, with arbitrary mass $M$. From (\ref{phase}) we see that their  wavefunctions are single-valued only when
\be
s = \tilde s
\ee
i.e. the left- and right winding numbers must be equal, which restricts to the non-spinning defects with $l_0 = \tilde l_0$.
For fermionic  probe particles, $J$ is half-integer and in a regular background the wavefunction should pick up a phase $-1$ when going around the origin.
From (\ref{phase}) we see that only the conical spaces with odd  $s$  appear regular to fermionic probes, which obey the familiar Neveu-Schwarz boundary conditions on the boundary cylinder.  The solutions with even $s$ do appear singular  to fermionic probes,
which experience the insertion of a  worldline defect  which causes them to obey Ramond boundary conditions on the boundary cylinder. For $s=0$ this reduces to  the familiar result that the zero-mass BTZ black hole lives in the Ramond sector of the dual CFT. Since we are in  2+1 dimensions we can also
consider probes with fractional spin, $J \in \NN + 1/n$, with $n$ a positive integer, whose wavefunctions should pick up a phase $e^{2 \p i\over n}$.
From (\ref{phase}) we see that only the conical spaces with $s = 1 {\rm \ mod \ } n$ appear regular to such probes.

To summarize, we have found that from the point of view of natural Chern-Simons observables all $(s, \tilde s)$ conical solutions appear regular in pure gravity, while adding bosonic probes selects the non-spinning solutions with $s = \tilde s$. Adding  fermionic or fractional spin matter further restricts the allowed values of $s$.

We end this section by commenting on the fate of the conical spaces when working in Euclidean rather than Lorentzian signature. In this case, the Chern-Simons gauge group is to be replaced by $SL(2,\CC )$, which has the property that all closed loops are contractible, $\p_1 (SL(2,\CC ))=0$. In this case, the analytic continuation of the conical solutions still yields smooth Euclidean solutions which however don't carry any topological winding charge. In fact, independent of the signature, the search for smooth solutions  which lie on the orbits of a constant covector  yields precisely the $(s,\tilde s)$ solutions and nothing else \cite{Castro:2011iw}.  We should note that there also exist coadjoint orbits which do not contain a constant covector \cite{Witten:1987ty}, whose role in 3D gravity remains to be fully understood (see \cite{Garbarz:2014kaa} for a  further discussion of these solutions).

\section{Quantizing the conical spaces}\label{secquant}
In the previous section we have provided some  plausibility arguments that the conical spaces  are  soliton-like solutions which appear smooth to observables in the Chern-Simons formulation,
and we would now like  to quantize them and identify them with states in a dual CFT. The standard method to quantize solitons is to make a perturbative expansion
of the  action in small  fluctuations around the soliton and proceed to quantize them. This gives the correction to the energy of the solution and the spectrum  of excited
 bound states  in an expansion in $\hbar$ (or, in our case, $1/c$) \cite{Coleman}.

Since the  Chern-Simons theory is topological all
on-shell fluctuations arise from gauge transformations (\ref{gaugetransf}) of $A, \tilde A$. However as we reviewed in section \ref{secas}, after imposing  asymptotically AdS boundary conditions the subset of transformations (\ref{asgauge}) should rather be seen as  generators of global symmetries which change  the solution. Therefore  the fluctuations we are to quantize come from acting on our solutions with asymptotic symmetries, i.e. from displacements on the coadjoint orbits of the solutions. In order to quantize them we use the fact
that  each coadjoint orbit carries a well-defined Poisson bracket, the Kirillov-Kostant bracket, under which the Poisson brackets of the Virasoro charges on the orbit satisfy the classical Virasoro algebra (\ref{VirPB}).

Upon quantizing  this Poisson bracket we expect to obtain  representations of the Virasoro group\footnote{See \cite{Kirillov} for a general perspective on coadjoint orbits and representation theory.} which we would like to determine\footnote{To be more precise, since our setup will be perturbative around the conical solutions, we will obtain representations of the Virasoro algebra rather than the Virasoro group.}  . As we shall see, the main obstacle  in carrying out this programme is that the  conical orbits have negative directions  for the energy  $l_0$. For this reason  it has so far proved impossible
to quantize these orbits by standard methods. We will here propose a semiclassical quantization of the conical orbits which leads to non-unitary highest weight Virasoro representations. The most important property of the resulting representation is that it will turn out to have a null vector at level $s$, and we will make the connection with Kac's classification of degenerate representations.

\subsection{Symplectic form on coadjoint orbits}
We start by briefly reviewing  the Kirillov-Kostant symplectic structure on the Virasoro coadjoint orbits, referring to \cite{Witten:1987ty}, \cite{Alekseev:1988ce} for more details. It will be convenient  to parameterize  points on the coadjoint orbit
of some fixed covector $(t(\f), c)$ by the finite group element  which maps $(t(\f), c)$ to the desired point\footnote{Once again we focus in this section on the left-moving sector, the right-moving sector being analogous.}.
In our case the group which acts on the orbits is  $Diff (S^1 )$, the group of  diffeomorphisms of the circle, i.e. periodic maps
\be
\f \mapsto F(\f ),\qquad F(\f + 2\p) = F(\f) + 2 \p.
\ee
The  infinitesimal coadjoint action (\ref{coadjinf}) integrates to the well-known finite transformation which involves the Schwarzian derivative $S(F)$:
\bea
(t(\f),  c) ) & \xlongrightarrow[]{F} & ( t_F(\f) , c )\\
  t_F(\f)&=&   t( F(\f)) (F')^2 - {c\over 24 \p} S (F)   \\
S(F) &\equiv & {F''' \over F'} - {3\over 2} \left({F'' \over F'}\right)^2
\eea
Combining this with (\ref{pairing}) we find the Virasoro charges along a coadjoint orbit
\be
l_n ((t_F ,c )) = \int_0^{2\p} e^{i n\f} \left( t(F(\f )) (F')^2 - {c\over 24 \p} S (F) \right). \label{chargesorb}
\ee
The Kirillov-Kostant symplectic form  at the point $( t_F , c )$ on the orbit of the covector $(t,c)$ can be written as \cite{Alekseev:1988ce}
\bea
\O &=& -\bigg \langle (t_F, c) , \left[ \left({\d F \over F'},0\right), \left({\d F \over F'},0\right)\right] \bigg  \rangle\\
&=&\int_0^{2 \p} d\f \left[ \left(t_0(F)  - {c \over 24 (F')^2} S(F) \right) \d F'\wedge \d F + {c \over 48\p} \left( {\d F \over F'} \right)'''\wedge  {\d F \over F'}\right]\label{sympl1}
\eea
where the brackets in the first line denote the pairing introduced in (\ref{pairing}). We have here introduced the notation $\d$ to denote the de Rham operator acting on  phase space variables, as oppposed to the spacetime de Rham operator $d$.
The main property of this symplectic form is that the Poisson brackets of the  charges along the orbit (\ref{chargesorb}) are guaranteed   to satisfy the classical Virasoro algebra (\ref{VirPB}).
Following \cite{Alekseev:1988ce}, (\ref{sympl1}) can be rewritten in a useful simplified  form as a total derivative
\be
\O =\int_0^{2 \p}d\f  \d \left( t(F(\f)) F' \d F -{c \over 48 \p} {\d F \over F'} \left( {F''' \over F'} - 2 \left({F'' \over F'}\right)^2\right) \right).\label{sympl}
\ee
\subsection{Conical spaces and exceptional orbits}\label{secflucts}
Next we want to apply these general formulas to the coadjoint orbits of the conical solutions.
We recall that the conical orbits are those of the constant covectors $(t_s,c)$, where
\be
t_s = - {c s^2 \over 48 \p}.
\ee
and $s > 1$.

Let us first analyze the symmetries of the conical orbits.
Any covector is trivially left invariant by $\hat c = (0,1)$, and  from (\ref{coadjinf}) we see that constant covectors are also invariant under the action of $l_0$. The conical covectors are  special
in that they are left invariant by two more generators; indeed one easily checks from (\ref{coadjinf}) that
\be
\d_{l_s} ((t_s,c)) =\d_{l_{-s}} ((t_s,c))=0.
\ee
 The infinitesimal transformations generated by the  charges $l_0, l_{\pm s}$\footnote{This  symmetry algebra  was referred to as a `twisted $sl(2,\RR )$' in \cite{Perlmutter:2012ds}.} integrate to finite reparameterizations of the circle of the form \cite{Balog:1997zz}
 \be
 F(\f) = {1\over i s} \ln {\a e^{i s \f} + \b \over \bar \b  e^{i s \f} + \bar \a}, \qquad |\a|^2 - |\b|^2 =1.
 \ee
These form an $s$-fold cover of $SO(2,1)$ which we  denote as $SO(2,1)^{(s)}$.
 Hence the conical orbits  have the  structure of the coset spaces $ Diff (S^1 ) / SO(2,1)^{(s)}$ and are often referred to as  exceptional coadjoint orbits.

Next we would like to compute the Virasoro charges along the conical orbits. We parameterize the elements of $Diff(S^1)$  as
\be
F(\f) = \f +  \sum_{n \in \ZZ} f_n e^{- in \f},
\ee
and reality of $F(\f)$ implies  that
\be
f_m^* = f_{-m}.\label{Freal}
\ee
 In view of the symmetries discussed above, the $f_n$ for $n \neq 0, s, -s$ are coordinates on the orbit  of $(t_s,c)$ (it will be a good check to see that the Virasoro charges are independent of $f_0, f_s$ and $f_{-s}$).
We start with the expression for the energy $l_0$, which using (\ref{chargesorb}) reads
\be
l_0 ((t_{s,F},c)) = -{c s^2 \over 24} + {c \over 12} \sum_{m \in \NN} m^2 (m^2 - s^2) |f_m|^2+ \calo (f^3 )\label{l0orb} \\
\ee
It's immediately clear that there is a major difference between the $s=1$ and $s>1$ cases: from (\ref{l0orb}) we see that  for $s=1$ the covector $(t_s,c)$ around which we are expanding is a minimum of the  energy, while for $s>1$ it is only a saddle point. The fluctuations $f_m$ with $|m|<s$ represent unstable directions of the conical spaces. Moreover, it can be shown that the energy on the conical orbits is unbounded below \cite{Witten:1987ty}.

The other Virasoro charges along the conical orbits are\footnote{The second expression requires some explanation. Applying (\ref{chargesorb}) one obtains
\be
l_{\pm s} ((t_{s,F},c)) = {c \over 24} \sum_{m \in \ZZ}  m (m \mp s) ( m^2 \pm ms - 3 s^2) f_m f_{\pm s - m}+ \calo (f^3 ).
\ee One can check this  expression does not depend on $f_0, f_s$ and $f_{-s}$, which can be made more explicit by adding zero in the form
\be
0 = \mp {c \over 24}\sum_{m \in \ZZ}  m (m \mp s) \left( m \mp {s \over 2}\right) s f_m f_{\pm s -m}
\ee
after which one obtains (\ref{Lsorb}).}
\bea
l_m ((t_{s,F},c)) &=& { i c \over 12} m(s^2-m^2) f_m + \calo (f^2 ); \qquad m \neq0,s,-s \\
l_{\pm s} ((t_{s,F},c)) &=& {c \over 24} \sum_{m \in \ZZ}  (m^2 - s^2) ( ( m\mp s)^2-s^2) f_m f_{\pm s - m} + \calo (f^3 ).\label{Lsorb}
\eea

Next we evaluate the expression  (\ref{sympl}) for the  symplectic form:
\be
\O = - {i c \over 12}\sum _{m \in \NN} m (m^2-s^2) \d f_m \wedge \d f_{-m} + \ldots.\label{symplpert}
\ee
where the dots mean that we have omitted terms of higher order in the $f_n$.
\subsection{Semiclassical expansion}
Now we would like to use the symplectic form (\ref{symplpert}) to quantize the conical orbits.
We start by introducing new coordinates $a_m, m \neq 0, s, -s,$  on the orbit:
\be
a_m= \sqrt{{ c\over 12}} f_m + \calo (f^2)\label{alphas}
\ee
in terms of which the symplectic form (\ref{sympl}) is
\be
\O = - i \sum_{m \in \NN} m(m^2- s^2) \d a_m\wedge \d a_{-m}\label{symplorb}.
\ee
Here, the $\calo (f^2)$ terms in (\ref{alphas}) are chosen such that (\ref{symplorb}) is exact without further higher order corrections;
it follows from Darboux's theorem\footnote{Or rather an infinite-dimensional generalization thereof explained in \cite{Witten:1987ty}, section 4.} that this is indeed possible.
 Note that the reality of $\O$ implies that
 \be a_m^* = a_{-m} ,\label{realas}\ee
 which at leading order order follows from (\ref{Freal}). The $a_m$ are related to canonical coordinates $x_m, p_m$, in terms of which $\O =  \sum_{m \in \NN_0 \backslash{ \{ s \} }} dp_m \wedge dx_m$, as follows:
\bea
a_{\pm m} &=& {1 \over \sqrt{2 |m(m^2-s^2)|}} ( x_m \mp i p_m), \qquad {\rm for \ } 0<m<s\\
a_{\pm m} &=& {1 \over \sqrt{2|m(m^2-s^2)|}} ( x_m \pm i p_m), \qquad {\rm for \ } m>s.
\eea
The Poisson brackets among the $a_m$ following from (\ref{symplorb}) are essentially those of harmonic oscillators,
\be
\{a_m, a_n \}_{PB} = {i  \d_{m,-n} \over m( m^2 - s^2)}\qquad {\rm for \ } m, n, \neq 0, \pm s\label{canPB}
\ee
If the Darboux-like coordinates (\ref{alphas}) and the expression for the Virasoro charges (\ref{Lsorb}) are known, we can in principle express the
Virasoro charges in terms of the $a_m$, yielding a perturbation series in powers of $1/c$.

Now we turn to the issue of quantizing the conical coadjoint orbits.  The $s=1$  case, which is the orbit of global AdS, can be quantized using standard methods \cite{Witten:1987ty}(it can be given a K\"ahler structure) and gives the unitary  representation based on the $SL(2,\RR )$  invariant vacuum. Hence we recover the standard result that global AdS is dual to the $SL(2,\RR )$  invariant vacuum in the CFT.

On the other hand, the $s>1$ orbits, corresponding to the conical excesses,  have so far defied quantization using standard methods \cite{Witten:1987ty}. This is most likely related to their being only saddle points of the energy:  we don't expect there to be any way of quantizing   such solutions in a way that  leads to  unitary highest weight Virasoro representations. Nevertheless,  we will   argue that they do give rise to a nonunitary highest weight Virasoro representation acting on a Fock space, which we will construct as a perturbation expansion in $1/c$. This, then, is what we will mean by `quantization' of these orbits.

We start by replacing  the classical coordinates $a_m$ on the orbit by quantum operators $A_m$ satisfying commutation relations obtained from (\ref{canPB})by  sending $-i \{\cdot, \cdot \}_{PB} \to [\cdot, \cdot ]$:
\be
[A_m, A_n] =  { \d_{m,-n} \over m (m^2- s^2)}\label{cancomm}.
\ee
From the classical expressions (\ref{Lsorb}) we can expect that  the quantum Virasoro generators, which we will denote by $L_m$, can be represented  as composite operators in terms of the oscillators $A_m$, with some as yet unknown ordering prescription. Let's start with the Virasoro energy $L_0$, which to our current level of accuracy can be written as
\be
L_0  = -{c s^2 \over 24}+ n_0 +  \sum_{m \in \NN} m^2 (m^2 - s^2) A_{-m} A_m + \calo ( c^{-1/2} ) \label{L0orb2}
\ee
where we have introduced an ordering constant  $n_0$ of order $c^0$ reflecting the ordering ambiguity  in the the quadratic term.
From its commutator with $A_{\pm m}$
 \be
 [L_0, A_{\pm m}] = \mp m  A_{\pm m} + \calo ( c^{-1/2} )
 \ee
 we see that, for large $c$, the $A_m$ with $m>0$  lower the $L_0$ eigenvalue, while  the $A_m$ with $m<0$  act as raising operators.
  Note that the oscillators with modes $|m| < s$ have a nonstandard minus sign in both the energy (\ref{L0orb2}) and the commutation relations (\ref{cancomm}) as a consequence of expanding around a saddle point.
  These modes are
similar to those arising  in the matter CFT of string theory from the timelike  field $X^0$ with negative kinetic energy.
As is the case for that theory,  we need to make some choice for  the `vacuum' state by stating which of the $A_m$ with $|m| < s$ annihilate it.
  No   choice is free from unpleasant features:  we must either give up having an  $L_0$ which is  bounded below or  having only positive norm states.
We will here make the choice to keep the energy levels positive by taking our the Fock vacuum $|0\rangle_s$ to be the state annihilated by the lowering operators,
\be
A_m |0\rangle_s =0 \qquad{\rm for \ } m>0 \label{defvac}.
\ee
By acting on  this vacuum with the raising operators, we build up a Fock space in the usual way. We require the inner product on this Fock space to satisfy \be A^\dagger_m= A_{-m}\ee reflecting the reality condition (\ref{realas}) and assume the ground state to be normalized as $\,_s\langle 0| 0 \rangle_s =1$.
 For $s>1$, our Fock space contains negative norm states,  since we must have $\| A_{-1}  |0\rangle_s \|^2=1/(1-s^2)<0$.
 The problems with normalizability are also evident from the expression for the position space wavefunction
\be
\Psi^{0}_{s} (x) \equiv \langle x| 0\rangle_s \sim e^{- \half \sum_{m \in \NN_0} {\rm sgn } (m-s) x_m^2}\label{vacwav}
\ee
i.e. the modes with $m<s$ come with  a wrong sign in the Gaussian wavefunction.

Let's discuss some properties of the expression for the Virasoro generators $L_m$ in terms of the oscillators $A_n$.
We will choose to represent the $L_m$  as creation-annihilation normal ordered expressions in the oscillators satisfying $L_m^\dagger = L_{-m}$, introducing unknown normal ordering constants where necessary. These constants are then to be fixed by requiring  that the $L_m$ obey the quantum Virasoro algebra
 \be
  [L_m,L_n] = (m-n) L_{m+n} + {c \over 12} m^3 \d_{m,-n}.
  \ee
  We will comment on a systematic procedure to achieve this below.
 Since the $L_m$  are normal ordered and of level $m$ in the oscillators, it follows that we will automatically have
\be
L_m |0\rangle_s =0 \qquad m>0
\ee
i.e.  our Fock space will furnish a highest weight (i.e. primary)  representation of the Virasoro algebra, which is however nonunitary because of the negative norm states mentioned above. In the following we will
determine precisely which nonunitary representations the conical spaces correspond to.

To the order at which we have been working, only $L_0$ requires the introduction of a normal ordering constant ($n_0$ in (\ref{L0orb2})), and the remaining Virasoro generators are given by
\bea
L_m &=&  - i \sqrt{c \over 12} m(m^2 - s^2) A_m  + \calo ( c^{0} )  \qquad {\rm for \ } m, \neq 0, \pm s\label{Lmorb2}\\
L_{\pm s} &=& \sum_{m\in \NN + \n_s} 2^{-\d_{m,0}} \left( m^2 - {s^2 \over 4} \right)  \left( m^2 - {9 s^2 \over 4}\right) A_{-m \pm {s \over 2}} A_{m \pm {s \over 2}}  + \calo ( c^{-1/2} )\label{Lsorb2}
\eea
where we have introduced a constant $\n_s$ which is 0 for $s$ even and $1/2$ for $s$ odd:
\be  \n_s \equiv { s\ {\rm mod}\ 2 \over 2} .\ee

Before turning to the determination of the normal ordering constant $n_0$, we outline our proposed  strategy for quantizing the conical orbits in a semiclassical expansion in $1/c$.
We have seen that Virasoro generators admit an expansion  in powers of $1/c$ in terms of the oscillators $A_m$ (given by our formulas (\ref{L0orb2},\ref{Lmorb2},\ref{Lsorb2}) plus corrections), where we consider the oscillators $A_m$ to be of  `order $c^0$' as the commutation
relations (\ref{cancomm}) are $c$-independent. We introduce the operators
$\calf_{m,n}$ measuring the failure of the Virasoro algebra to hold:
\be
\calf_{m,n}  = [L_m,L_n] - (m-n) L_{m+n} - {c \over 12} m^3 \d_{m,-n}.
\ee
When expressed in terms of the oscillators $A_m$ these admit an expansion in powers of $1/c$:
\be
\calf_{m,n} = \sum_{\a \in \NN /2} { \calf_{m,n}^{( \a-2)}\over c^{\a- 2}} .
\ee
with $ \calf_{m,n}^{(\a)}$ independent of $c$.
We expect  that, by adjusting a finite number of normal ordering constants up to some order $(1/c)^\b$, it will possible to make the
 Virasoro commutation relations hold to order $(1/c)^\b$ in the sense that
 \be
 \calf_{m,n}^{(\a)} =0\qquad {\rm for}\ -2 \leq \a \leq \b.
 \ee
Proceeding in this way we  build up an oscillator realization of the Virasoro algebra, order by order in $1/c$. Although we don't have a general proof that this procedure is consistent, we will see that it gives sensible and satisfying results at the first nontrivial order.

It is instructive to check the vanishing of the first few   $\calf_{m,n}^{(\a)}$  using (\ref{L0orb2},\ref{Lmorb2},\ref{Lsorb2}) and the commutation relations (\ref{cancomm}). The operators $\calf_{m,n}^{(-2)}$ and
$\calf_{m,n}^{(-3/2)}$ vanish trivially as they are given by commutators with the  c-number term in $L_0$. For the operators $\calf_{m,n}^{(-1)}$, $\calf_{m,0}^{(-1/2)}$ and $\calf_{m,s}^{(-1/2)}$ one finds
\bea
\calf_{m,n}^{(-1)} &=& -{1 \over 12} m (m^2 -s^2) n (n^2 -s^2) [A_m, A_n] - {m \over 12} (m^2 - s^2)\d_{m,-n} \nonu
\calf_{m,0}^{(-1/2)} &=& -{i \over \sqrt{12}} m (m^2 -s^2) \sum_{n \in \NN} n^2 (n^2-s^2) [A_m, A_{-n} A_{n}] + {i m^2(m^2-s^2) \over \sqrt{12}} A_m\nonu
\calf_{m,s}^{(-1/2)} &=&  -{i \over \sqrt{12}} m (m^2 -s^2) \sum_{n \in \NN+ \n_s}  2^{-\d_{n,0}} \left( n^2 - {s^2 \over 4} \right)  \left( n^2 - {9 s^2 \over 4}\right) [A_m,  A_{-n + {s \over 2}} A_{n + {s \over 2}}  ] \nonu
&&+{i \over \sqrt{12}}(m^2-s^2)((m+s)^2 - s^2)  A_{m+s}  .
\eea
These operators do not involve any normal ordering constants and hence their vanishing is guaranteed because the classical Virasoro algebra holds by construction. It is straightforward to verify that they  indeed vanish upon using (\ref{cancomm}) and the commutator
\be
[ A_m,  A_{-n + {s \over 2}} A_{n + {s \over 2}}] = {\d_{m,n- {s \over 2}} + \d_{m,- n- {s \over 2}}\over m(m^2-s^2)} A_{m+ s} 
\ee
which follows from (\ref{cancomm}). 

\subsection{One-loop energy correction}
The first nontrivial check on our proposed method to quantize the conical orbits comes at order $c^0$, where the operator $\calf_{s,-s}^{(0)}$ involves the normal ordering constant $n_0$ in (\ref{L0orb2}).
Since $c$ multiplies the action (\ref{SCS}), $1/c$ is the loop-counting parameter in our theory and the constant $n_0$, which is of order $c^0$, can be interpreted as the 1-loop correction to the
energy of the conical solutions.
The quickest way to determine $n_0$ is to impose that the expectation value of $\calf_{s,-s}^{(0)}$  in the Fock vacuum vanishes:
\be
\,_s\langle 0 | \calf_{s,-s}^{(0)} |0\rangle_s =0.
\ee
This leads to
\be
n_0 = {1 \over 2 s} \,_s\langle 0 | [ L_s',   L_{-s}' ]|0\rangle_s.\label{novac}
\ee
where $ L_s'$ ( $L_{-s}'$ ) is the piece of $L_s$ ($L_{-s}$) in (\ref{Lsorb2}) which involves two annihilation (creation) operators:
\be
L_{\pm s}'= \sum_{m= \n_s}^{s/2 -1} 2^{-\d_m,0} \left( m^2 - {s^2 \over 4} \right)  \left( m^2 - {9 s^2 \over 4}\right) A_{-m \pm {s \over 2}} A_{m \pm {s \over 2}}.\label{Lprime}
\ee
From this expression we can evaluate (\ref{novac}) using (\ref{cancomm}) and obtain
\bea
n_0 &=& \sum_{m= \n_s}^{s/2 -1} 2^{-\d_m,0} \left( {9 s \over 8} - {m^2 \over 2 s}\right)\\
&=& {1 \over 24} (s-1) (1 + 13 s)\label{n0}.
\eea

It remains to check that, with the  value (\ref{n0}) of $n_0$, the operator $\calf_{s,-s}^{(0)}$ indeed vanishes. It is given by
\bea
\calf_{s,-s}^{(0)} &=& \sum_{m,n \in \NN + \n_s}\left( 2^{-\d_{m,0}-\d_{n,0}} \left( m^2 - {s^2 \over 4} \right)  \left( m^2 - {9 s^2 \over 4}\right) \left( n^2 - {s^2 \over 4} \right)  \left( n^2 - {9 s^2 \over 4}\right)\times \right.\nonu
&&\left.[ A_{-m + {s \over 2}} A_{m + {s \over 2}}, A_{-n - {s \over 2}} A_{n - {s \over 2}}]\right) -2 s n_0 - 2 s \sum_{m \in \NN} m^2 (m^2 - s^2) A_{-m} A_m
\eea
and can be shown to vanish upon using 
\begin{gather}
[ A_{-m + {s \over 2}} A_{m + {s \over 2}}, A_{-n - {s \over 2}} A_{n - {s \over 2}}]=
 \left( { A_{-m+ {s\over 2}} A_{m- {s\over 2}} \over (m + {s \over 2}) \left( (m+{s \over 2})^2 - s^2 \right) }
  + { A_{-m- {s\over 2}} A_{m+ {s\over 2}} \over (-m + {s \over 2}) \left( (m-{s \over 2})^2 - s^2 \right) } \right)  \d_{m,n}\nonu+ \left({ A_{m+ {s\over 2}} A_{-m- {s\over 2}} \over (-m + {s \over 2}) \left( (m-{s \over 2})^2 - s^2 \right) }+
{ A_{m- {s\over 2}} A_{-m+ {s\over 2}} \over (m + {s \over 2}) \left( (m+{s \over 2})^2 - s^2 \right) }  \right)\d_{m,-n}.
\end{gather}
Similarly one verifies  that  $\calf_{s,0}^{(0)}$ is  zero as well. To check the vanishing of the remaining order $c^0$ operators $\calf_{m,n}^{(0)}$ would require going to the next order  in our expansion (\ref{Lmorb2}) and would involve determining further integration constants, which we will leave for  further study.

Summarizing, we have computed the Virasoro  energy of the conical spaces to 1-loop order to be
\be
 \boxed{\,_s\langle 0 | L_0 |0\rangle_s = -{s^2 c \over 24}  + {1 \over 24} (s-1) (1 + 13 s)+ \calo ( c^{-1} ).}\label{1loopen}
\ee
\subsection{Null states and Kac's degenerate  representations}
We have seen in section \ref{secflucts} that the coadjoint orbits of the conical solutions are rather special in that they have a three parameter isotropy group $SO(2,1)^{(s)}$. As a consequence, the classical Virasoro charges  $l_0,l_s,l_{-s}$ on the coadjoint orbit can be expressed in terms of the other $l_m$, see (\ref{Lsorb}). It is natural to expect that also
on the quantum level  the corresponding Virasoro representation is such that $L_0,L_s,L_{-s}$ can be expressed in terms of the remaining $L_m$ \cite{Witten:1987ty}.
This property is the hallmark of a Virasoro representation which contains a null vector at level $s$:
 acting with the vanishing combination of $L_{-s}$ and the other generators  on the Fock vacuum  we obtain a null vector, and conversely, in  a highest weight representation with a null vector at level $s$ it is possible to express $L_0, L_{\pm s}$ as a function of the remaining $L_m$ (see \cite{Witten:1987ty} for a proof).

For definiteness, from (\ref{Lsorb2}) we see that, to our level of accuracy, we can express $L_0,L_s,L_{-s}$ as
 \bea
L_0 &=& - {c s^2 \over 24} + n_0+  {12 \over c}\sum_{m\in \NN \backslash \{0,s\}} {L_{-m} L_m \over m^2 - s^2} + \calo ( c^{-3/2} ) \\
L_{\pm s} &=& {12\over c} \sum_{ \begin{array}{cc} \scriptstyle{ m \in \NN + \n_s} \\ \scriptstyle{ m\neq s/2, 3s/2} \end{array}}{2^{-\d_{m,0}}  L_{-m\pm s/2} L_{m\pm s/2} \over m^2- {s^2 \over 4} }
+ \calo (c^{-3/2} )\label{LsitoLm}
\eea
The null vector obtained by acting with (\ref{LsitoLm}) on the Fock vacuum is, after a shift in the summation variable and a reordering of terms:
\be
\boxed{\left( L_{-s} - {6 \over c}  \sum_{ m = 1}^{s -1} { L_{-m} L_{m-s} \over m(m-s) } + \calo (c^{-3/2} )\right) |0\rangle_s=0 \label{nullvectors}}
\ee
This expression  is the operator version
of a classical formula derived in \cite{Witten:1987ty}. The second term is  again (minus) the operator $L_{-s}'$ defined in (\ref{Lprime}). 

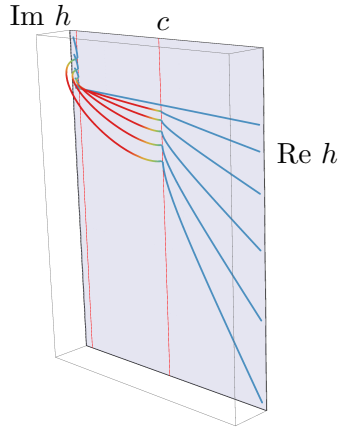
\begin{figure}
\begin{center}
\begin{picture}(100,150)
\put(0,0){\includegraphics[height=150pt]{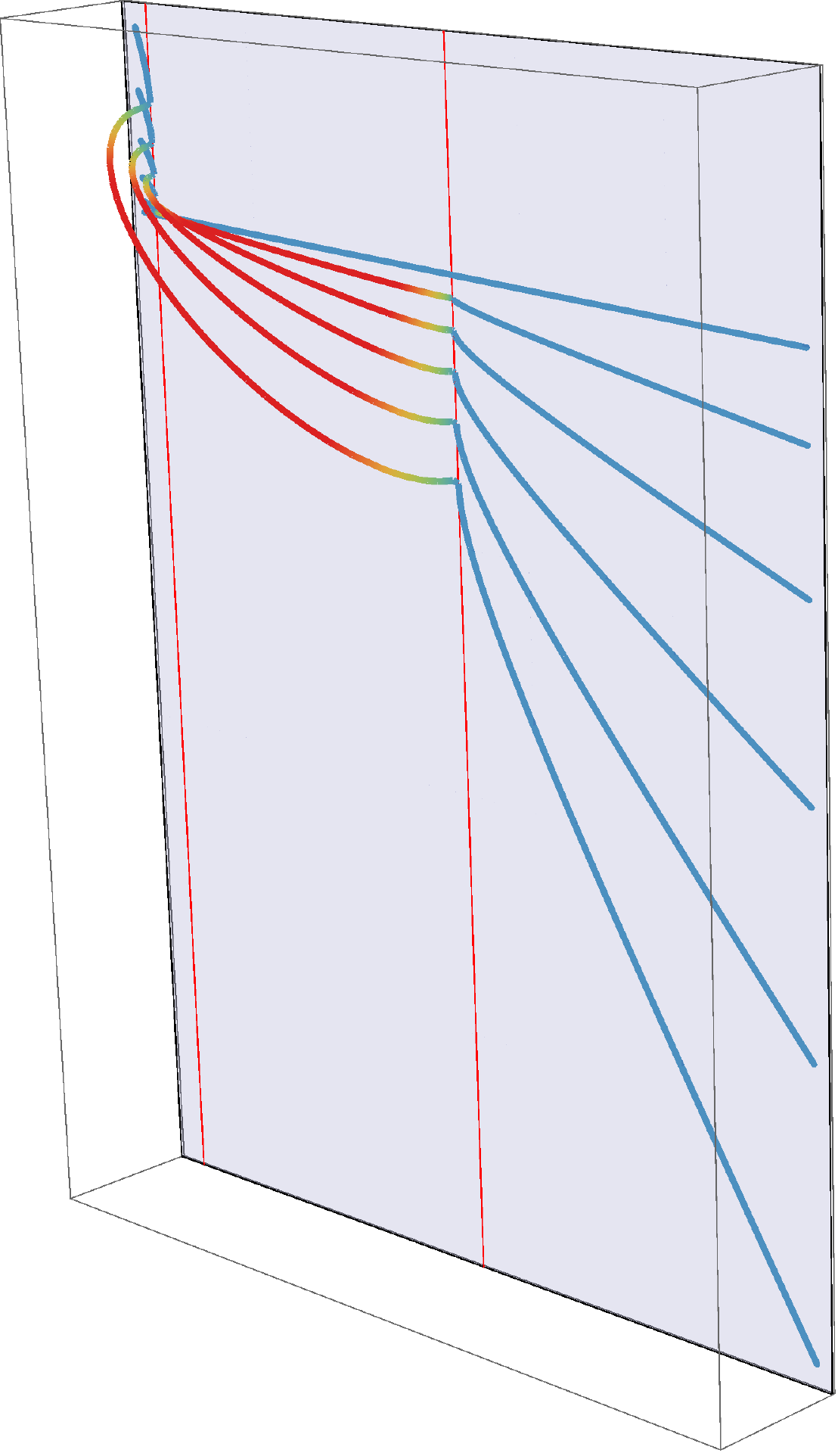}}
\put(45,150){$c$}
\put(90,100){Re $h$}
\put(-10,152){Im $h$}
\end{picture}\end{center}
\caption{The conformal weights $h_{(1,s)}$ for $1 \leq s \leq 6$ as a function of the central charge $c$. The red lines correspond to $c=1$ and $c=25$, between which the conformal weights become complex.}\label{1sfig}
\end{figure}

Virasoro representations containing a null vector are referred to as degenerate and  were
classified by Kac \cite{Kac:1978ge} (see also \cite{Feigin:1981st}). We will now identify precisely which Kac representations  the quantized conical spaces correspond to.  Kac's result can be summarized as follows: for any value of the central
charge $c$, there is a degenerate representation for every pair of nonzero natural numbers\footnote{For small values of the central charge some of those representations can become equivalent, but this doesn't play a role in the large $c$ regime we are considering.} $(r,s)$. It contains a null vector at level $r s$ and is based on a primary of weight
\be
h_{(r,s)} = -\frac{1}{24} -\frac{r s}{2}+\frac{1}{48} (13-c) \left(r^2+s^2\right) +\frac{1}{48} \sqrt{(1-c)(25-c)} \left(r^2-s^2\right).
\ee
The degenerate representations are usually only considered at small values of the central charge, $0<c<1$, since only in that regime they have a conformal weights  above the vacuum and stand a chance of being unitary. However, they exist as nonunitary representations also for $c$ outside this range. Note that, for $r\neq s$, $h_{(r,s)}$ becomes complex in the range $1<c<25$ and is real again for $c\geq 25$.
Comparing the large $c$ expansion
\be
h_{(r,s)}  \sim -\frac{ s^2 c}{24}+\frac{1}{24} \left(-12 r s+13 s^2-1\right)+ \calo (c^{-1} )
\ee
with (\ref{1loopen}) leads to the unambiguous identification of the quantized orbit of the  conical space labelled by $s$   with the degenerate representation of type $(1,s)$ of the left-moving Virasoro algebra. The conformal weight of the representations $h_{(1,s)}$ is plotted as a function of $c$ in figure \ref{1sfig}.
 Similarly, the right-moving winding number $\tilde s$ corresponds to the
 representation of type $(1,\tilde s)$ of the right-moving Virasoro algebra.

A further check comes from  comparing the null vectors: the explicit form of the null vector at level $s$ of the $(1,s)$ degenerate  representation was first
derived by Benoit and Saint-Aubin \cite{Benoit:1988aw}. Their expression (see (2.7) in \cite{Benoit:1988aw}) is written as a sum over partitions of $s$, and the leading contributions at large $c$ come from
the partitions with one and two elements. It is straightforward to check that  these contributions coincide with  our expression (\ref{nullvectors}).
This identification between the conical states and $(1,s)$  degenerate representations was already proposed in \cite{Perlmutter:2012ds} based on classical considerations  in the bulk, and we have hereby confirmed  agreement also on the 1-loop level.

\section{Discussion}

We end with some comments and directions for future work.
\begin{itemize}
\item We argued that the conical spaces  play a role  in nonunitary versions of holography and are dual to primaries
of degenerate Virasoro representations. The origin of the nonunitary behaviour came from expanding around a saddle point of the Virasoro energy.
This seems to be the generic situation in the known consistent   nonunitary quantum field theories. For example in Vafa's  nonunitary holographic theories  \cite{Vafa:2014iua}, which are  based on large $N$ gauge theories where the gauge group is a supergroup,  the classical ground state in the super matrix model description also has negative energy directions. We also want to remark that our  Fock vacua $|0\rangle_s$  seem rather similar  to
 the Kodama state in 4D quantum gravity \cite{Kodama:1990sc}, which is  a saddle point of the Hamiltonian as was pointed out in  \cite{Witten:2003mb}. Our results suggest that this state may yet play a role in examples of nonunitary holography.
\item We computed the 1-loop energy correction for conical spaces from the bulk perspective using operator methods, by quantizing the natural Poisson bracket on coadjoint orbits.  It would be instructive to rederive this result from a 1-loop determinant in the path integral formalism. Here the challenge is to derive the action governing the boundary graviton fluctuations. A natural guess for this  action is the so-called geometric action \cite{Alekseev:1988ce} which reproduces the Poisson bracket on the coadjoint orbits by construction, and which should be obtainable from the original Chern-Simons action (some evidence for this appears in \cite{Alekseev:1988ce},\cite{Bershadsky:1989mf}).
    \item The method outlined in section \ref{secquant} allows one to derive a perturbative expression for the null vector in the degenerate representation in a large $c$-expansion.
The expression for  the exact null vectors \cite{Benoit:1988aw} can similarly be derived  from a large $c$ expansion \cite{Bauer:1991ai}, but the relation between the two approaches is as yet unclear.  Clarifying this link would be especially interesting since the knowledge of the exact null vector implies knowledge of the energy correction to  all orders.
\item We provided evidence that the Chern-Simons  formulation of 3D gravity accommodates the degenerate Virasoro representations of the type $(1,s)$. While a CFT containing only these primaries does have a spectrum which is closed under OPE's, our experience with minimal model CFTs at $c<1$ suggests that such a theory cannot be modular invariant and that this requires the inclusion of  the more general  $(r,s)$ representations. It therefore natural to ask what type of matter has  to be added to pure gravity
     in order to get all $(r,s)$ representations in the spectrum. There is evidence \cite{Perlmutter:2012ds} that the theory that accomplishes this  is the Prokushkin-Vasiliev theory \cite{Prokushkin:1998bq} at the value $\l =2$ of the vacuum parameter, where it reduces to gravity coupled to a somewhat unusual scalar field, whose excitations around the conical spaces give the $(r,s)$ representations.
      In fact, our computation of the energy shift (\ref{1loopen}) was the missing ingredient in the matching of the bulk 1-loop partition function
       to the partition function of such a CFT at large $c$. It is an interesting open question whether the full  CFT partition function including $1/c$ corrections is (presumably in some formal sense) modular invariant, and whether it can be reproduced from the bulk side.
\item The degenerate Virasoro representations display various interesting features at small values of the central charge ($0<c<1$), which would be very interesting
 to understand physically as strong coupling phenomena in the bulk\footnote{As we reviewed in section \ref{secCS}, with our choice of gauge group $c$ is quantized in units of 24, so to consider small values of $c$ we should either consider a covering group or work in Euclidean signature, where $c$ is not quantized.}.
 Examples are the fact that some of the  $(r,s)$ representations  can become equivalent to each other and  become unitary at special values of the central charge in the strong coupling regime $0<c<1$. For those special values of $c$ a truncation of the spectrum of  $(r,s)$ representations gives a unitary modular invariant theory, leading to the Virasoro minimal models, some of which may be interpretable as strongly coupled gravity theories \cite{Castro:2011zq}.
    \item As already mentioned in the Introduction, the conical solutions in  pure gravity which we studied in this paper are the more tractable cousins of similar solutions\footnote{The interpretation as topological solitons characterized by a winding number seems to be special to the pure gravity case, since the higher spin theories are based on the gauge groups
        $SL(N,\RR )$ (for the Lorenzian theory)  or $SL(N,\CC )$ (for the Euclidean theory), which have $\p_1\left(SL(N,\RR )\right) = \ZZ_2$ and $\p_1\left(SL(N,\CC )\right) = 1$.
        }  \cite{Castro:2011iw} in
    higher spin theories with asymptotic $W$-symmetry \cite{Henneaux:2010xg},\cite{Campoleoni:2010zq}. It would be interesting
       to extend the regularity arguments of section \ref{seccon} using point particle probes to the higher spin case, as well to understand whether they, too, are  singular in the  metric-like formulation of the theory \cite{Campoleoni:2012hp}. For the higher spin conical solutions, a similar holographic interpretation as degenerate nonunitary representations of the $W$-algebra was proposed in \cite{Perlmutter:2012ds}. The arguments for this identifications were purely classical, and it would be interesting to generalize our method to compute quantum corrections to these solutions.
       While  the concepts of finite $W$-symmetry transformations  and coadjoint orbits are certainly much less understood than  their
    Virasoro counterparts, we do feel that it should be possible to extend the perturbative setup of the current work the higher spin case.
\end{itemize}

\section*{Acknowledgements}
 I would like to  thank Andrea Campoleoni, Frederik Denef, Rajesh Gopakumar, Gustavo Lucena G\'{o}mez and Tom\'{a}\v{s} Proch\'{a}zka  for useful discussions.  This research was supported by the Grant Agency of the Czech Republic under the grant
14-31689S. I also want to acknowledge   support from the collaborative grant WBI/14-1 for a visit to ULB Brussels  which sparked some of the questions that led to this work.

\end{document}